\documentclass{rjparticle}
\usepackage{graphicx}

\title{Mass formulas for single-charm tetraquarks with Fermi-Breit hyperfine interaction}
\author[]{V. Borka Jovanovi\'{c}}
\author[]{D. Borka}
\affil[]{Atomic Physics Laboratory (040), Vin\v{c}a Institute of Nuclear Sciences, University of Belgrade, P.O. Box 522, 11001 Belgrade, Serbia
\\Email:{\em vborka@vinca.rs}, {\em dusborka@vinca.rs}}


\begin{document}
\maketitle
\begin{abstract}
In this paper we present the main results of our investigation of
the $cq \bar{q} \bar{q}$ single-charm scalar tetraquarks and their
SU(3)$_\mathrm{F}$ representations: $\overline{15}_S$, $\bar{3}_S$,
$6_A$ and $\bar{3}_A$. We use the Fermi-Breit interaction
Hamiltonian with SU(3) flavor symmetry breaking to determine the
masses of the single-charm tetraquarks. We also discuss mass spectra
obtained from meson and baryon mass fits. The mass spectra are very
similar to those obtained with Glozman-Riska hyperfine interaction,
and they indicate that some of the experimentally detected states
may have tetraquark nature.
\end{abstract}

\section{Introduction}

The possible existence of four-quark states for light flavor
dimensions, as well as some predictions for tetraquark spectroscopy,
was first suggested by Jaffe \cite{jaff77a}. In Ref. \cite{jaff77b}
it is also provided a framework for a quark-model classification of
the many two-quark-two-antiquark states. In Ref. \cite{silv92} the
energies of diquonia $q^2(\bar q^2)$ with orbital angular momentum
$L$ = 0 are calculated and compared to the treshold energies. The
first observations of the scalar charmed mesons have been reported
\cite{aube03,mika04,abe04,link04,evdo04}. In Refs.
\cite{beve03,beve04,barn04} some observed mesons (as for instance
D$_s^+$(2317)) are explained as a scalar $c \bar{s}$ systems. Many
studies appeared in the past on D$_s^+$(2317)
(\cite{bard03,wang06,wang07} and references therein). Possible
interpretations include tetraquark states or molecular $DK$ or
$DK^*$ states \cite{swan06}. The origin of the lightest scalar
mesons, in particular, the light $\sigma$-meson is given in
\cite{molo09} in the framework of the instanton liquid model of the
QCD vacuum. Terasaki and Hayashigaki
 \cite{haya06,tera06} have investigated the decay rates of the
members of the same multiplet in class of four-quark mesons. Also,
in Refs. \cite{tera05,dmit05,niel06} it is shown that the existence
of some exotic states with the four-quark structures might be
expected. There are more results indicating that a
diquak-antidiquark structure is acceptable for some observed states
\cite{nico04,liu04,brac05,brit05,thoo08}. In Refs.
\cite{vija06,vija07} a mixture of conventional quark-antiquark
states and four-quark components is considered. Also, there are
predictions for some hidden charm states to be tetraquarks. For
example, in \cite{eber08} the masses of the ground state heavy
tetraquarks are calculated in the framework of the relativistic
quark model, based on the quasipotential approach in quantum
chromodynamics. These authors found that some exotic meson
candidates can be tetraquark states with hidden charm. A concise
overview of mesons with heavy quarks including charmed mesons and
charmonium (or charmonium-like) states is given in \cite{zhu08}. The
mass spectrum of the scalar hidden charm and bottom tetraquark
states is studied in \cite{wang09}.

In this paper we perform a schematic study (two quark interaction)
of the masses of the single-charm $cq \bar{q} \bar{q}$ tetraquarks
in the SU(3) flavor representations. We consider states with the
spin-parity quantum numbers $J^P$ = 0$^+$ ($J$ = total angular
momentum, $P$ = parity). Using the colored version of the
Fermi-Breit (FB) hyperfine interaction (HFI)
\cite{ruju75,libe77,fran81,luch91,silv92} we investigate the
possible four-quark structure of these mesons. According to the mass
constraint, experimentally indicated states can contain only one
$c$-quark. If each of them contains no heavy quarks, mass is too
low, and if it contains one heavy quark ($b$ or $t$) or two
$c$-quarks mass will be too high. Only two possible solutions are:
$cq \bar{q} \bar{q}$ that we investigate and $\bar{c} \bar{q} qq$
that is analyzed by Liu et al. \cite{liu04}.

We showed \cite{bork10} that the constituent quark masses are very
sensitive to the system in which they are contained, and their
values differ less or more in different systems. That is why we
analyzed tetraquark masses using independent meson and baryon fits
for constituent quark masses. These fits satisfy Feynman-Hellmann
theorem for FB interaction \cite{ronc95a,ronc95b}. Constituent quark
masses in tetraquarks are somewhat different for both cases: quark
masses from meson fit and from baryon fit \cite{bork10,giam95}.

We choose relatively simple FB HFI because it nicely satisfies
Roncaglia inequalities for mass differences \cite{ronc95a,ronc95b}.
We deal with the open charm states. Since our intention is to
investigate the possible tetraquark nature of 27 $cq \bar{q}
\bar{q}$ states, it is not necessary to develop an advanced
relativistic quark model. That is why masses of the scalar charmed
tetraquarks are discussed in the framework of the nonrelativistic
quark model, in which the mass of a hadron is considered to be the
sum of the constituent quark masses and contributions of the FB HFI.
As it will be shown later, our simple model is quite sufficient for
investigating wave functions and masses of these states. Also, this
model is used to get estimates of the theoretical meson and baryon
masses. We prefer to deal with four quarks instead of diquarks
because our system under consideration consists of one heavy and
three light quarks.

In this paper we present detailed calculation of masses using FB
interaction, and it is the first time this interaction is applied to
27 states of $cq \bar{q} \bar{q}$ tetraquarks in order to derive
formulas for masses. We also compare these results with another
phenomenological interaction: Glozman-Riska hyperfine interaction
(GR HFI) \cite{gloz96}.

\section{Analysis and Method}

We discussed single-charm tetraquarks in Ref. \cite{bork07}, where
the flavor wave functions and masses of $c q \bar{q} \bar{q}$
tetraquarks are calculated using GR HFI. Now we present detailed
calculation of masses of $c q \bar{q} \bar{q}$ tetraquarks using FB
HFI.

We analyze the tetraquark states with one charm quark which is
singlet under the transformation of SU(3)$_\mathrm{F}$. There are
four multiplets according to product: $3 \otimes \bar 3 \otimes \bar
3 \otimes 1 = \overline {15}_S + \bar 3_S + 6_A + \bar 3_A$, i.e.
two anti-triplets, one anti-15-plet and one sextet. Young diagrams
for these SU(3)$_\mathrm{F}$ multiplets, as well as the weight
diagrams, can be found in Figs. 1 - 5 of \cite{bork07}. In these
tetraquark multiplets, all states have a charm number equal to 1.
Labels for all 27 states are the same as in Refs.
\cite{bork07,bork08}. These labels are taken only by analogy with
baryons, but of course they are not baryons. Many authors use the
same labeling \cite{tera05,dmit05}. The strong FB HFI Hamiltonian
\cite{silv92} may be written in the following form:
\begingroup
\setlength{\abovedisplayskip}{0pt}
\setlength{\belowdisplayskip}{0pt}
\begin{equation}
H_{FB} = C\sum\limits_{i > j = 1}^4 {\frac{\vec \sigma_i \vec
\sigma_j}{m_i m_j}\left( {\lambda_i^C \lambda_j^C} \right)},
\label{equ01}
\end{equation}
\endgroup

\noindent were $\sigma_i$ are the Pauli spin matrices, $\lambda^C_i$
are the color Gell-Mann matrices and $C$ is a constant proportional
to strong hyperfine structure constant $\alpha_c$. Hamiltonian
(\ref{equ01}) has explicit color and spin exchange dependence and
implicit (by way of quark masses) flavor dependence. Its
contribution to tetraquark masses is:
\begingroup
\setlength{\abovedisplayskip}{0pt}
\setlength{\belowdisplayskip}{0pt}
\begin{equation}
m_{\nu,FB} = \left\langle \nu \right|\left\langle \chi
\right|H_{FB} \left| \chi \right\rangle \left| \nu \right\rangle,
\label{equ02}
\end{equation}
\endgroup

\noindent where $\chi$ denotes the spin wave function and $\nu$ -
flavor wave function. For total masses $m_{\nu}$ we have:
\begingroup
\setlength{\abovedisplayskip}{0pt}
\setlength{\belowdisplayskip}{0pt}
\begin{equation}
m_{\nu} = m_{\nu,0} + m_\mathrm{\nu,FB},
\label{equ03}
\end{equation}
\endgroup

\noindent where $m_{\nu,0}$ are masses without influence of FB HFI.

Here we have to mention that the mixing of states is taken into
account. In Table I of the paper \cite{bork07} the four-quark
content and quantum numbers of scalar $cq \bar{q} \bar{q}$
tetraquarks are given. One can see from that table which states mix
due to the same quantum numbers. So we took into account the mixing
when calculating masses of these tetraquarks.

\section{Results and Discussion}

In our model (total spin = 0), the corresponding symmetric $\chi_S$
and antisymmetric $\chi_A$ spin functions have the following forms:
\begingroup
\setlength{\abovedisplayskip}{0pt}
\setlength{\belowdisplayskip}{0pt}
\begin{equation}
\left| {\chi_S } \right\rangle = \frac{-1}{2 \sqrt 3}\left|
{\uparrow \downarrow \uparrow \downarrow + \downarrow \uparrow
\uparrow \downarrow + \uparrow \downarrow \downarrow \uparrow +
\downarrow \uparrow \downarrow \uparrow - 2 \uparrow \uparrow
\downarrow \downarrow -2 \downarrow  \downarrow \uparrow \uparrow }
\right\rangle,
\label{equ04}
\end{equation}

\begin{equation}
\left| {\chi_A } \right\rangle = \frac{1}{2}\left| {\uparrow
\downarrow \uparrow \downarrow - \downarrow \uparrow \uparrow
\downarrow - \uparrow \downarrow \downarrow \uparrow + \downarrow
\uparrow \downarrow \uparrow } \right\rangle.
\label{equ05}
\end{equation}
\endgroup

The spin wave functions are symmetric ($\chi_S$) or antisymmetric
($\chi_A$) under interchange between the pair of quarks and between
the pair of antiquarks (not for quark-antiquark interchange).
Tetraquarks are bosons, i.e. they have integer spin (for scalar
tetraquarks it equals 0). For the open charm system, $cq \bar{q}
\bar{q}$, the interaction between the light quarks $q$ and the $c$
quark is suppressed in the heavy quark limit \cite{yasu07}. Thus, in
the first approximation, three light quarks are decoupled from the
heavy quark and there can be considered states compounded by $u$,
$d$ and $s$ as triquarks or color nonsinglet baryons (in the bound
state) \cite{yasu07}. Symmetry of their total wave functions is
determined as in the case of fermions because there are three quarks
in SU(3) group: $q \bar{q} \bar{q}$, where $q$ = $u$, $d$, $s$. In
the case of fermions the total wave function has to be
antisymmetric, as well as the color state function in the case of
all hadrons, and therefore the particles from multiplets
$\overline{15}_S$ and $\bar{3}_S$ have the symmetric spin and flavor
wave functions ($\chi_S, \nu_S$), while the particles from
multiplets $6_A$ and $\bar{3}_A$ have the antisymmetric spin and
flavor wave functions ($\chi_A, \nu_A$).

The calculation of FB contribution to tetraquark masses will be
described in more details. First we will explain how to calculate
the $\vec \sigma_i \vec \sigma_j$ products of Pauli spin matrices
for each pair in the scalar system of four quarks. These products
are the expected values of spin matrix elements. We use the spin
operator eigenvalues for triplet and singlet states: (+1) and (-3),
respectively. We also use the following values of the matrix
elements $\vec \sigma_1 \vec \sigma_2$:
\begingroup
\setlength{\abovedisplayskip}{0pt}
\setlength{\belowdisplayskip}{0pt}
\begin{equation}
\begin{array}{c}
\left\langle { \uparrow \downarrow } \right|{\vec \sigma_1}{\vec \sigma_2}\left| { \uparrow \downarrow } \right\rangle = - 1 = \left\langle { \downarrow \uparrow } \right|{\vec \sigma_1}{\vec \sigma_2}\left| { \downarrow \uparrow } \right\rangle \\
\left\langle { \uparrow \uparrow } \right|{\vec \sigma_1}{\vec \sigma_2}\left| { \uparrow \uparrow } \right\rangle = 1 = \left\langle { \downarrow \downarrow } \right|{\vec \sigma_1}{\vec \sigma_2}\left| { \downarrow \downarrow } \right\rangle \\
\left\langle { \downarrow \uparrow } \right|{\vec \sigma_1}{\vec \sigma_2}\left| { \uparrow \downarrow } \right\rangle = 2 = \left\langle { \uparrow \downarrow } \right|{\vec \sigma_1}{\vec \sigma_2}\left| { \downarrow \uparrow } \right\rangle.
\end{array}
\label{equ06}
\end{equation}
\endgroup

\noindent In addition to that, for total spin $S$ and its projection
$m_s$, for symmetric state it holds:
\begingroup
\setlength{\abovedisplayskip}{0pt}
\setlength{\belowdisplayskip}{0pt}
\begin{equation}
\begin{array}{c}
\left| {S = 1,m_s = 1} \right\rangle = \left| { \uparrow
\uparrow } \right\rangle \\
\left| {S = 1,m_s = 0} \right\rangle = \frac{1}{\sqrt 2}\left| {
\uparrow \downarrow + \downarrow \uparrow } \right\rangle \\
\left| {S = 1,m_s = - 1} \right\rangle = \left| { \downarrow
\downarrow } \right\rangle,
\end{array}
\label{equ07}
\end{equation}
\endgroup

\noindent from where we get $\left| { \uparrow  \downarrow +
\downarrow \uparrow } \right\rangle = \sqrt 2 \left| {S = 1,m_s = 0}
\right\rangle$.

\noindent For antisymmetric state it holds:
\begingroup
\setlength{\abovedisplayskip}{0pt}
\setlength{\belowdisplayskip}{0pt}
\begin{equation}
\left| {S = 0,m_s = 0} \right\rangle = \frac{1}{\sqrt 2}\left| {
\uparrow \downarrow - \downarrow \uparrow } \right\rangle,
\label{equ08}
\end{equation}
\endgroup

\noindent and combining expressions (\ref{equ07}) and (\ref{equ08}),
we get: $\left| {\uparrow \downarrow } \right\rangle =
\frac{1}{\sqrt 2}\left( {\left| {S = 1} \right\rangle + \left| {S =
0} \right\rangle } \right)$ and $\left| {\downarrow \uparrow }
\right\rangle = \frac{1}{\sqrt 2}\left( {\left| {S = 1}
\right\rangle - \left| {S = 0} \right\rangle } \right)$.

Then we calculate symmetric $\left\langle {\chi_S} \right|{\vec
\sigma_i}{\vec \sigma_j}\left| {\chi_S} \right\rangle $ and
antisymmetric matrix elements $\left\langle {\chi_A} \right|{\vec
\sigma_i}{\vec \sigma_j}\left| {\chi_A} \right\rangle $ for the
following pairs: $q_1 q_2$, $q_1 \bar q_3$, $q_1 \bar q_4$, $q_2
\bar q_3$, $q_2 \bar q_4$, $\bar q_3 \bar q_4$, and apply them to
relations for spin wave functions (\ref{equ04}) and (\ref{equ05}).
For example, for $q_1 q_2$ pair, we use only spins of the first and
second particle, therefore the other two spins e.g. of the third and
forth particle are used only to determine which addends are
non-zero. In case of symmetric spin wave function it leads from eq.
(\ref{equ09}) to eq. (\ref{equ10}):
\begingroup
\setlength{\abovedisplayskip}{0pt}
\setlength{\belowdisplayskip}{0pt}
\begin{equation}
\begin{array}{l}
\left\langle \chi_S \right| \vec \sigma_1 \vec \sigma_2 \left| \chi_S \right\rangle = \\
= \frac{-1}{2 \sqrt 3}\left\langle {\left( {\uparrow_1 \downarrow_2 + \downarrow_1 \uparrow_2} \right) \uparrow_3 \downarrow_4 + \left( {\uparrow_1 \downarrow_2 + \downarrow_1 \uparrow_2} \right) \downarrow_3 \uparrow_4} \right. + \\
~~~~~~~~~~~ \left. { -2 \uparrow_1 \uparrow_2 \downarrow_3 \downarrow_4 - 2 \downarrow_1 \downarrow_2 \uparrow_3 \uparrow_4} \right| \times \\
\times (\vec \sigma_1 \vec \sigma_2) \left| {\left( {\uparrow_1 \downarrow_2 + \downarrow_1 \uparrow_2} \right) \uparrow_3 \downarrow_4 + \left( {\uparrow_1 \downarrow_2 + \downarrow_1 \uparrow_2} \right)\downarrow_3 \uparrow_4 + } \right. \\
~~~~~~~~~~~~ \left. { -2 \uparrow_1 \uparrow_2 \downarrow_3 \downarrow_4 - 2 \downarrow_1 \downarrow_2 \uparrow_3 \uparrow_4} \right\rangle \frac{-1}{2 \sqrt 3},
\end{array}
\label{equ09}
\end{equation}

\begin{equation}
\begin{array}{l}
\left\langle \chi_S \right| \vec \sigma_1 \vec \sigma_2 \left| \chi_S \right\rangle = \\
= \frac{1}{4 \cdot 3}\left( \right.\left\langle \right.\underbrace {\left( {\uparrow \downarrow + \downarrow \uparrow} \right)}_I + \underbrace {\left( {\uparrow \downarrow + \downarrow \uparrow} \right)}_{II} - 2 \uparrow \uparrow - 2 \downarrow \downarrow \left| \right. \times \\
~~\times (\vec \sigma_1 \vec \sigma_2)\left. {\left| {\left( {\uparrow \downarrow + \downarrow \uparrow} \right) + \left( {\uparrow \downarrow + \downarrow \uparrow} \right) - 2 \uparrow \uparrow - 2 \downarrow \downarrow} \right\rangle } \right).
\end{array}
\label{equ10}
\end{equation}
\endgroup

\noindent Nevertheless, we see that addends labeled with I and II in
eq. (\ref{equ10}) cannot be combined into non-zero matrix elements
because the spins of the third and fourth particles are $\uparrow
\downarrow$ in case I, and $\downarrow \uparrow$ in case II. So we
derive:
\begingroup
\setlength{\abovedisplayskip}{0pt}
\setlength{\belowdisplayskip}{0pt}
\begin{equation}
\begin{array}{l}
\left\langle \chi_S \right| \vec \sigma_1 \vec \sigma_2 \left| \chi_S \right\rangle = \\
\frac{1}{12}\left( {\sqrt 2 \left\langle {S = 1,m_s = 0} \right|(\vec \sigma_1 \vec \sigma_2)\left| {S = 1,m_s = 0} \right\rangle \sqrt 2 + } \right. \\
+ \sqrt 2 \left\langle {S = 1,m_s = 0} \right|(\vec \sigma_1 \vec \sigma_2)\left| {S = 1,{m_s} = 0} \right\rangle \sqrt 2 + \\
+ \left( -2 \right)\left\langle {S = 1,m_s = 1} \right|(\vec \sigma_1 \vec \sigma_2)\left| {S = 1,m_s = 1} \right\rangle \left( -2 \right) + \\
\left. { + \left( -2 \right)\left\langle {S = 1,m_s = -1} \right|(\vec \sigma_1 \vec \sigma_2)\left| {S = 1,m_s = - 1} \right\rangle \left( - 2 \right)} \right),
\end{array}
\label{equ11}
\end{equation}
\endgroup

\noindent from where we obtain this result: $\left\langle \chi_S
\right|\vec \sigma_1 \vec \sigma_2\left| \chi_S \right\rangle =
\frac{1}{12}\left( {2 \cdot 1 + 2 \cdot 1 + 4 \cdot 1 + 4 \cdot 1}
\right) = 1$.

In this way we get the results for all quark pairs $q_i q_j$:
\begingroup
\setlength{\abovedisplayskip}{0pt}
\setlength{\belowdisplayskip}{0pt}
\begin{equation}
\begin{array}{c}
\left\langle \chi_S \right| \vec \sigma_1 \vec \sigma_2 \left| \chi_S \right\rangle = 1 = \left\langle \chi_S \right| \vec \sigma_3 \vec \sigma_4 \left| \chi_S \right\rangle \\
\left\langle \chi_S \right| \vec \sigma_1 \vec \sigma_3 \left| \chi_S \right\rangle = -2 = \left\langle \chi_S \right| \vec \sigma_2 \vec \sigma_4 \left| \chi_S \right\rangle \\
\left\langle \chi_A \right| \vec \sigma_1 \vec \sigma_2 \left| \chi_A \right\rangle = -3 = \left\langle \chi_A \right| \vec \sigma_3 \vec \sigma_4 \left| \chi_A \right\rangle \\
\left\langle \chi_A \right| \vec \sigma_1 \vec \sigma_3 \left| \chi_A \right\rangle = 0 = \left\langle \chi_A \right| \vec \sigma_2 \vec \sigma_4 \left| \chi_A \right\rangle. \\
\end{array}
\label{equ12}
\end{equation}
\endgroup

\begin{table}[ht!]
\centering
\caption{The products $\lambda_i \lambda_j$ (Gell-Mann
matrices for color SU(3)$_\mathrm{C}$) and the products $\vec
\sigma_i \vec \sigma_j$ (Pauli spin matrices) for symmetric and
antisymmetric multiplets.}
\begin{tabular}{lcc}
\noalign{\smallskip}
\hline
multiplet & $q_i q_j$, $\bar{q}_i \bar{q}_j$ & $q_i \bar{q}_j$ \\
\noalign{\smallskip}
\hline
\noalign{\smallskip}
\raisebox{-0.10ex}[0cm][0cm]{$\overline{15}_S$ and $\bar{3}_S$} & $\lambda_1 \lambda_2 = \lambda_3 \lambda_4 =
-\frac{8}{3}$ & $\lambda_1 \lambda_3 = \lambda_1 \lambda_4 = \lambda_2 \lambda_3 = \lambda_2 \lambda_4 =
-\frac{4}{3}$ \\
& $\vec \sigma_1 \vec \sigma_2 = \vec \sigma_3 \vec \sigma_4 = 1$ & $\vec \sigma_1 \vec \sigma_3 = \vec \sigma_1 \vec \sigma_4 = \vec \sigma_2 \vec \sigma_3 = \vec \sigma_2 \vec \sigma_4 =
-2$ \\
\raisebox{-0.10ex}[0cm][0cm]{$6_A$ and $\bar{3}_A$} & $\lambda_1 \lambda_2 = \lambda_3 \lambda_4 =
-\frac{8}{3}$ & $\lambda_1 \lambda_3 = \lambda_1 \lambda_4 = \lambda_2 \lambda_3 = \lambda_2 \lambda_4 =
-\frac{4}{3}$ \\
& $\vec \sigma_1 \vec \sigma_2 = \vec \sigma_3 \vec \sigma_4 = -3$ & $\vec \sigma_1 \vec \sigma_3 = \vec \sigma_1 \vec \sigma_4 = \vec \sigma_2 \vec \sigma_3 = \vec \sigma_2 \vec \sigma_4 =
0$ \\
\noalign{\smallskip}
\hline
\end{tabular}
\label{tab01}
\end{table}

\begin{table}[ht!]
\centering
\caption{Masses of scalar $cq \bar{q} \bar{q}$
tetraquarks distributed in SU(3)$_\mathrm{F}$ multiplets, with
mixing between states with the same quantum numbers.
$m_\mathrm{\nu,0}$ are tetraquark masses without influence of FB HFI
and $m_\mathrm{\nu,FB}$ are FB HFI contributions to tetraquark
masses.}
\begin{tabular}{llcc}
\noalign{\smallskip}
\hline
multiplet & $cq \bar{q} \bar{q}$ & $m_{\nu,0}$ ($m_{u}$ =
$m_d$)&
$m_\mathrm{\nu,FB}$ ($m_u$ = $m_d$) \\
\noalign{\smallskip}
\hline
\noalign{\smallskip}
\raisebox{-0.10ex}[0cm][0cm]{$\overline{15}_S$} & $\Xi$ & $m_u$ +
2$m_s$ + $m_c$ &
$ {\frac{8}{3}}C \left({{\frac{2}{m_s m_c}} + {\frac{2}{m_u m_s}} - {\frac{1}{m_u m_c}} - {\frac{1}{m_s^2}}} \right)$ \\
& $\Sigma_s$ & 2$m_u$+$m_s$ + $m_c$ & $ {\frac{8}{3}}C \left({{\frac{1}{m_u^2}} + {\frac{1}{m_s m_c}}} \right)$ \\
& $\Delta$ & 3$m_u$ + $m_c$& $ {\frac{8}{3}}C \left({{\frac{1}{m_u^2}} + {\frac{1}{m_u m_c}}} \right)$ \\
& $\Sigma$ & 2$m_u$ + $m_s$ + $m_c$&
$ {\frac{8}{3}}C \left({{\frac{2}{m_u m_c}} + {\frac{2}{m_u m_s}} - {\frac{1}{m_s m_c}} - {\frac{1}{m_u^2}}} \right)$ \\
& & & \\
\raisebox{-0.10ex}[0cm][0cm] {$\overline{15}_S$ -- $\bar{3}_S$}
& D$_s$ & 2$m_u$ + $m_s$ + $m_c$; 3$m_s$ + $m_c$ & $ {\frac{8}{3}}C
\left({{\frac{1}{m_u^2}} + {\frac{1}{m_s m_c} }} \right)$;
$ {\frac{8}{3}}C \left( {{\frac{1}{m_s^2}} + {\frac{1}{m_s m_c} }} \right) $ \\
& D & 3$m_u$ + $m_c$; $m_u$ + 2$m_s$ + $m_c$ & $ {\frac{8}{3}}C
\left({\frac{1}{m_u^2}} + {\frac{1}{m_u m_c}} \right)$;
${\frac{4}{3}}C \left({{\frac{3}{m_s^{2}}} + {\frac{2}{m_u m_c}} - {\frac{1}{m_u^2} }} \right) $ \\
& & & \\
\raisebox{-0.10ex}[0cm][0cm]{$6_A$}
& $\Sigma_s$ & 2$m_u$ + $m_s$ + $m_c$& $ 8C \left( {\frac{1}{m_u m_s}} + {\frac{1}{m_u m_c}} \right)$ \\
& $\Omega$ & 2$m_u$ + $m_s$ + $m_c$& $ 8C \left( {\frac{1}{m_u^2}} + {\frac{1}{m_s m_c}} \right)$ \\
& & & \\
\raisebox{-0.10ex}[0cm][0cm]{$\bar{3}_A$}
& D$_s$ & 2$m_u$ + $m_s$ + $m_c$& $ 8C \left( {\frac{1}{m_u m_s}} + {\frac{1}{m_u m_c}} \right)$ \\
& & & \\
\raisebox{-0.10ex}[0cm][0cm] {$6_A$ -- $\bar{3}_A$} &
D & 3$m_u$ + $m_c$; $m_u$ + 2$m_s$ + $m_c$ & $ 8C \left(
{\frac{1}{m_u m_s}} + {\frac{1}{m_s m_c}} \right)$;
$ 8C \left( {\frac{1}{m_u^2}} + {\frac{1}{m_u m_c}} \right)$ \\
\noalign{\smallskip}
\hline
\end{tabular}
\label{tab02}
\end{table}

In Table \ref{tab01} we give the products of Gell-Mann matrices for
color SU(3)$_\mathrm{C}$ and the products of Pauli spin matrices.
When we put values for $\vec \sigma_i \vec \sigma_j$ and $\lambda_i
\lambda_j$ from Table \ref{tab01} into equation (\ref{equ02}), we
get the following FB HFI contributions for symmetric and
antisymmetric multiplets:
\begingroup
\setlength{\abovedisplayskip}{0pt}
\setlength{\belowdisplayskip}{0pt}
\begin{eqnarray}
m_{\nu,FB,S} = -\frac{8}{3}C \left\langle {\nu_S} \right| \frac{1}{m_1 m_2} + \frac{1}{m_3 m_4} \left| {\nu_S} \right\rangle + ~~~~~~~~~~\nonumber \\
+ \frac{8}{3}C \left\langle {\nu_S} \right| \frac{1}{m_1 m_3} + \frac{1}{m_1 m_4} + \frac{1}{m_2 m_3} + \frac{1}{m_2 m_4} \left| {\nu_S} \right\rangle,
\label{equ13}
\end{eqnarray}
\noindent for symmetric multiplets $\overline{15}_S$ and
$\bar{3}_S$, and
\begin{equation}
m_{\nu,FB,A} = 8C\left\langle {\nu_A} \right| {\frac{1}{m_1 m_2} + \frac{1}{m_3 m_4}} \left| {\nu_A} \right\rangle,
\label{equ14}
\end{equation}
\endgroup

\noindent for antisymmetric multiplets $6_A$ and $\bar{3}_A$. The
flavor wave functions $\nu_S$ and $\nu_A$ for the scalar $cq \bar{q}
\bar{q}$ tetraquarks are given in Table II of Ref. \cite{bork07}.
The masses of tetraquarks predicted from our model are given in
Table \ref{tab02}. The mixing between states with the same quantum
numbers is included. There is mixing between states from symmetric
multiplets $\overline{15}_S$ and $\bar{3}_S$, and also between
antisymmetric multiplets $6_A$ and $\bar{3}_A$, while symmetric and
antisymmetric multiplets do not mix with each other. We first show
our predictions for spectra when FB HFI is not included (the third
column of the table) and then we show the FB HFI influence (the
forth column).

Here we study mass spectra of single-charm tetraquarks using FB HFI
in sche\-matic approximation (two-particle interaction). The masses
of constituent quarks $m_u$ (= $m_d$), $m_s$ and $m_c$ and the
constant are calculated from $\chi^2$ fitting of the equations for
meson and for baryon masses, with FB interaction included, to the
experimental meson and baryon masses \cite{bork10}. As one can see
from Table 2 in Ref. \cite{feng08a} or Table 2 in Ref.
\cite{feng08b} and from references therein, our predictions for
constituent quark masses are similar to masses obtained using
different phenomenological models. We use masses for these mesons:
light pseudoscalar mesons $\pi$, $K$, light vector mesons $\rho$,
$K^{*}$, $\omega$, $\varphi$, charmed mesons $D$, $D^*$ and strange
charmed mesons $D_S$, $D_S^*$. We did not calculate $\eta$ and
$\eta'$ contribution because of their mixing and because they cannot
be described within such a model. Also, we use masses for these
baryons: light baryon octet N, $\Sigma$, $\Xi$, $\Lambda$, light
baryon decuplet $\Delta$, $\Sigma^{*}$, $\Xi^{*}$, $\Omega$ and
heavy baryons $\Sigma_\mathrm{c}$, $\Lambda_\mathrm{c}$,
$\Sigma_\mathrm{c}^{*}$, $\Omega_\mathrm{c}$. For each set of
equations, the minimized $\chi^{2}$ values for masses are calculated
by formula (14) given in Ref. \cite{bork07}. The corresponding
experimental masses are taken from the "Particle Data Group" site:
http://pdg.lbl.gov \cite{PDG10}.

From the $\chi^{2}$ fit of all meson masses (when mesons with two
$c$-quarks are excluded) we obtained the following values
\cite{bork10} (Table III): $m_u$ = $m_d$ = 314.75 MeV, $m_s$ =
466.80 MeV, $m_c$ = 1627.31 MeV and the constant $C^m$ = $1.5546
\times 10^7 \mathrm{MeV}^3$. These values for quark masses here we
use for calculating the tetraquark masses. For the constant $C$ we
use value $C^{tetra}$ which is different from $C^{meson}$. We fitted
$C^{tetra}$ in that way to obtain the mass of the lowest state from
$\bar{3}_A$ equal as the D$_s^+$(2317) meson: $C^{tetra}$ = $-5.80
\times 10^6 \mathrm{MeV}^3$. From the fit of all baryon masses (when
barions with two $c$-quarks are excluded), in \cite{bork10} (Table
III) we obtained: $m_u$ = $m_d$ = 365.69 MeV, $m_s$ = 530.08 MeV,
$m_c$ = 1700.17 MeV and the constant $C^b$ = $1.2513 \times 10^7
\mathrm{MeV}^3$. Here, we calculated $C^{tetra}$ = $-11.90 \times
10^6 \mathrm{MeV}^3$ (in that way to obtain the mass of the lowest
state from $\bar{3}_A$ equal as the D$_s^+$(2317) meson).

\begin{table}[ht!]
\centering
\caption{The values of theoretical masses $m_m$ (in MeV)
of some mesons with FB HFI included, when we use constituent quark
masses and the constant $C^m$ obtained from the meson fit (Table III
of Ref. \cite{bork10}). $m_m^{exp}$ are experimental masses
\cite{PDG10} and $\Delta m_m$ is the absolute difference between
these two values.}
\begin{tabular}{crrr}
\noalign{\smallskip}
\hline
meson & $m_m$ (MeV) & $m_m^{exp}$ (MeV) & $\Delta m_m$ (MeV) \\
\noalign{\smallskip}
\hline
\noalign{\smallskip}
$\pi$ & 159 & 140 & 19 \\
$K$ & 464 & 494 & 30 \\
$\rho$ & 786 & 776 & 10 \\
$K^*$ & 887 & 892 & 5 \\
$\omega$ & 786 & 783 & 3 \\
$\varphi$ & 1005 & 1020 & 15 \\
$D$ & 1851 & 1869 & 18 \\
$D^*$ & 1972 & 2010 & 38 \\
$D_S$ & 2033 & 1968 & 65 \\
$D_S^*$ & 2015 & 2012 & 3 \\
\noalign{\smallskip}
\hline
\end{tabular}
\label{tab03}
\end{table}

\begin{table}[ht!]
\centering
\caption{The values of theoretical masses $m_b$ (in MeV)
of some baryons with FB HFI included, when we use constituent quark
masses and the constant $C^b$ obtained from the baryon fit (Table
III of Ref. \cite{bork10}). $m_b^{exp}$ are experimental masses
\cite{PDG10} and $\Delta m_b$ is the absolute difference between
these two values.}
\begin{tabular}{crrr}
\noalign{\smallskip}
\hline
baryon & $m_b$ (MeV) & $m_b^{exp}$ (MeV) & $\Delta m_b$ (MeV) \\
\noalign{\smallskip}
\hline
\noalign{\smallskip}
$N$ & 957 & 940 & 17 \\
$\Sigma$ & 1179 & 1190 & 11 \\
$\Xi$ & 1319 & 1315 & 4 \\
$\Lambda$ & 1121 & 1116 & 5 \\
$\Delta$ & 1237 & 1232 & 5 \\
$\Sigma^{*}$ & 1373 & 1385 & 12 \\
$\Xi^{*}$ & 1513 & 1530 & 17 \\
$\Omega$ & 1657 & 1672 & 15 \\
$\Sigma_\mathrm{c}$ & 2438 & 2455 & 17 \\
$\Lambda_\mathrm{c}$ & 2291 & 2285 & 6 \\
\noalign{\smallskip}
\hline
\end{tabular}
\label{tab04}
\end{table}

The values of theoretical masses of some mesons and baryons with FB
HFI included, using constituent quark masses and the constants
$C^m$, $C^b$ given in Table III of Ref. \cite{bork10} (the upper
rows which correspond to FB HFI) we present in Tables \ref{tab03}
and \ref{tab04}. If we compare Tables \ref{tab03} and \ref{tab04} we
can notice that theoretically obtained masses of baryons using FB
HFI are in better agreement with the corresponding experimental
masses than masses of mesons.

The values of masses (in MeV) of scalar $cq \bar{q} \bar{q}$
tetraquarks, obtained from the meson fit, are given in Table
\ref{tab05}, and their masses obtained from the baryon fit are given
in Table \ref{tab06}. These tables with tetraquark masses show us
where the 27 tetraquark states are expected. There are uncertainties
in calculating masses because we use the model with schematic
interaction.

\begin{table}[ht!]
\centering
\caption{The values of masses (in MeV) of scalar $cq
\bar{q} \bar{q}$ tetraquarks distributed in SU(3)$_\mathrm{F}$
multiplets, with mixing between states with the same quantum
numbers, obtained from the meson fit. $m_\mathrm{\nu,0}$ (MeV) are
tetraquark masses without influence of FB HFI, $m_\mathrm{\nu,FB}$
(MeV) are FB HFI contributions to tetraquark masses and
m$_\mathrm{\nu}$ (MeV) are the total tetraquark masses.}
\begin{tabular}{llccc}
\noalign{\smallskip}
\hline
multiplet & tetraquark & $m_\mathrm{\nu,0}$ (MeV)& $m_\mathrm{\nu,FB}$ (MeV) & $m_\mathrm{\nu} $ (MeV) \\
\noalign{\smallskip}
\hline
\noalign{\smallskip}
\raisebox{-0.10ex}[0cm][0cm]{$\overline{15}_S$}& $\Xi$ & 2876& -150& 2726 \\
& $\Sigma_s$& 2724& -176& 2547 \\
& $\Delta$ & 2572& -186& 2385 \\
& $\Sigma$ & 2724& -94& 2629 \\
& & & & \\
\raisebox{-0.10ex}[0cm][0cm]{$\overline{15}_S$ -- $\bar{3}_S$ } &
D$_s$($\overline{15}_S$ -- $\bar{3}_S$)& 2724; 3028& -176; -91& 2547; 2936 \\
& D($\overline{15}_S$ -- $\bar{3}_S$)& 2572; 2876& -186; -59& 2385; 2817 \\
& & & & \\
\raisebox{-0.10ex}[0cm][0cm]{$6_A$}& $\Sigma_s$ & 2724& -406& 2317 \\
& $\Omega$ & 2724& -529& 2194 \\
& & & & \\
$\bar{3}_A$& D$_s$& 2724& -406& 2317 \\
& & & & \\
$6_A$ -- $\bar{3}_A$ &
D($6_A$ -- $\bar{3}_A$)& 2572; 2876& -377; -559& 2195; 2317 \\
\noalign{\smallskip}
\hline
\end{tabular}
\label{tab05}
\end{table}

\begin{table}[ht!]
\centering
\caption{The values of masses (in MeV) of scalar $cq
\bar{q} \bar{q}$ tetraquarks distributed in SU(3)$_\mathrm{F}$
multiplets, with mixing between states with the same quantum
numbers, obtained from the baryon fit. $m_\mathrm{\nu,0}$ (MeV) are
tetraquark masses without influence of FB HFI, $m_\mathrm{\nu,FB}$
(MeV) are FB HFI contributions to tetraquark masses and
m$_\mathrm{\nu}$ (MeV) are the total tetraquark masses.}
\begin{tabular}{llccc}
\hline
multiplet & tetraquark & $m_\mathrm{\nu,0}$ (MeV)& $m_\mathrm{\nu,FB}$ (MeV) & $m_\mathrm{\nu} $ (MeV) \\
\noalign{\smallskip}
\hline
\noalign{\smallskip}
\raisebox{-0.10ex}[0cm][0cm]{$\overline{15}_S$}& $\Xi$ & 3126& -234& 2892 \\
& $\Sigma_s$& 2962& -273& 2689 \\
& $\Delta$ & 2797& -288& 2509 \\
& $\Sigma$ & 2962& -157& 2805 \\
& & & & \\
\raisebox{-0.10ex}[0cm][0cm]{$\overline{15}_S$ -- $\bar{3}_S$ } &
D$_s$($\overline{15}_S$ -- $\bar{3}_S$)& 2962; 3290& -273; -148& 2689; 3142 \\
& D($\overline{15}_S$ -- $\bar{3}_S$)& 2797; 3126& -288; -102& 2509; 3024 \\
& & & & \\
\raisebox{-0.10ex}[0cm][0cm]{$6_A$}& $\Sigma_s$ & 2962& -644& 2317 \\
& $\Omega$ & 2962& -818& 2144 \\
& & & & \\
$\bar{3}_A$& D$_s$& 2962& -644& 2317 \\
& & & & \\
$6_A$ -- $\bar{3}_A$ &
D($6_A$ -- $\bar{3}_A$)& 2797; 3126& -597; -865& 2200; 2261 \\
\noalign{\smallskip}
\hline
\end{tabular}
\label{tab06}
\end{table}

\begin{figure}[ht!]
\centering
\includegraphics[width=0.53\textwidth]{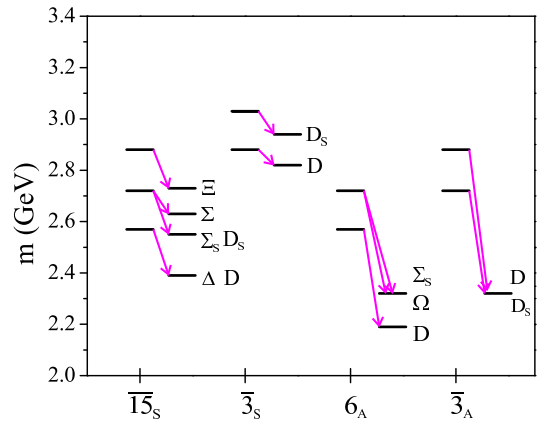}
\caption{Tetraquark mass spectrum from the meson fit without (left
column) and with (right column) FB HFI, both with SU(3)$_\mathrm{F}$
symmetry breaking.}
\label{fig01}
\end{figure}

\begin{figure}[ht!]
\centering
\includegraphics[width=0.53\textwidth]{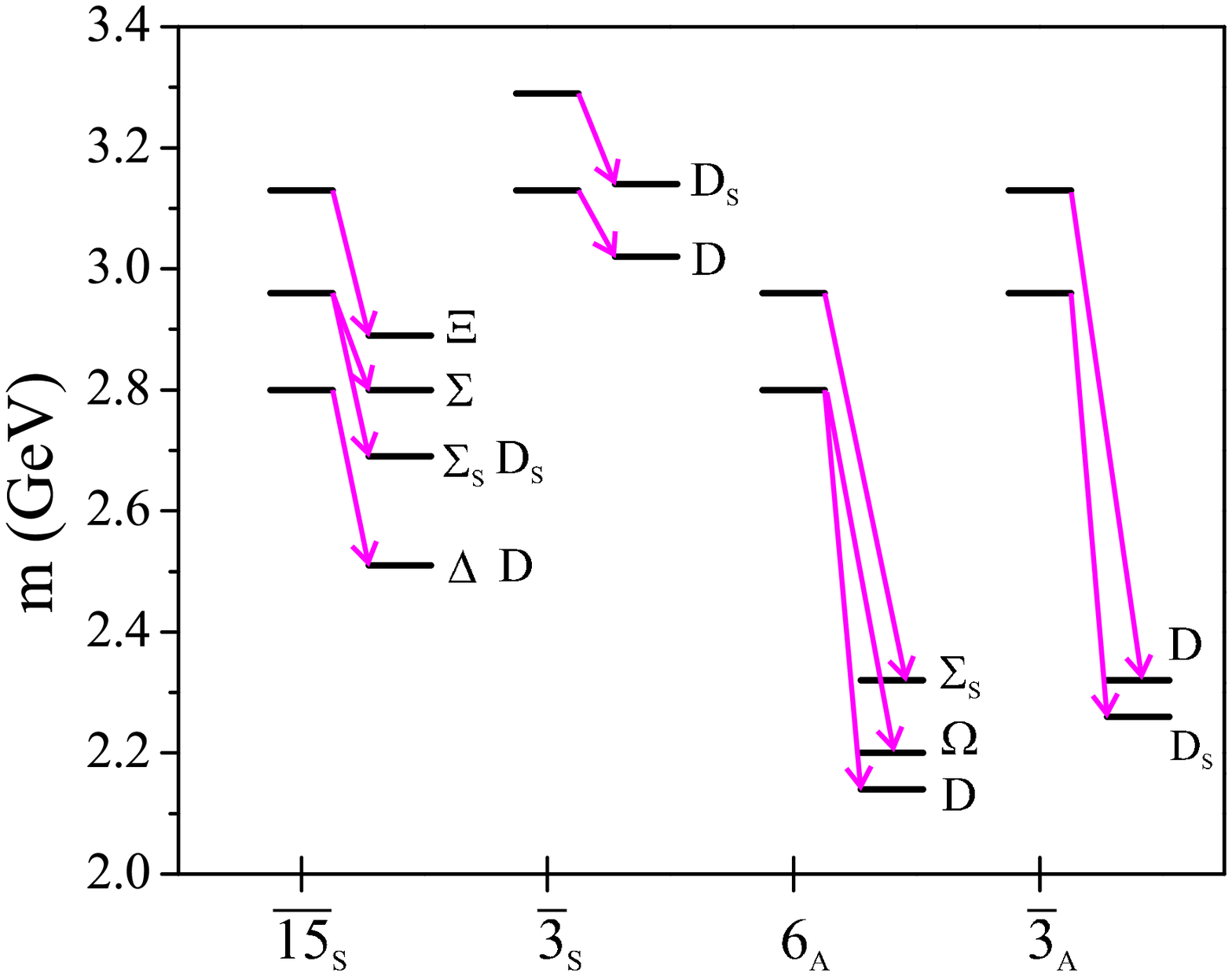}
\caption{the same as Fig. 1, but for baryon fit.}
\label{fig02}
\end{figure}

The tetraquark mass spectra are given in Figs. \ref{fig01} and
\ref{fig02} (labels are the same as in Ref. \cite{bork07}). It can
be noticed that HFI determines mass splitting in the spectrum, i.e.
its fine structure. And for mixing it can be said that it separates
two states. When comparing Fig. \ref{fig01} (or Fig. \ref{fig02})
from this paper with Fig. 6 from \cite{bork07} and Fig. 1 from
\cite{bork08}, it can be noticed that GR HFI reduces masses (except
for $\overline{15}_S$ -- $\bar{3_S}$ mixed states) more than FB HFI,
but that difference is not so significant. One can see that FB HFI
reduces the obtained masses and causes splitting between $\Sigma$
and $\Sigma_s$ (D$_s$) in $\overline{15}_S$ and between $\Sigma_s$
and $\Omega$ in $6_A$, and GR HFI only causes splitting between
$\Sigma_s$ and $\Omega$. Also, there is difference for
$\overline{15}_S - \bar{3}_S$ mixed states when comparing with GR
HFI, but the forms of tetraquark spectra with FB and GR interactions
are similar. It is interesting to note that the spectra are similar
although the one interaction is color-spin, and the other one is
flavor-spin.

From spectrum it is possible to identify D$_s^+$(2317) as the lowest
state in multiplet $\bar{3}_A$. It agrees with the identification of
D$_s^+$(2317) with tetraquark state which was considered in Ref.
\cite{maia05}. In Ref. \cite{maia05}, for D$_s^+$(2632) it was
claimed to be a candidate for a tetraquark D$_s$, but they proposed
different total angular momentum ($J^p$ = 2$^+$). In the present
paper, the authors consider scalar tetraquark only. Also, from
spectrum we can see that FB HFI reduces the obtained masses for all
states.

States D$_s^+$(2632) \cite{evdo04} and D$^0$(2308) \cite{abe04} are
mixed states: D$_s^+$(2632) is from $\overline{15}_S$ -- $\bar{3}_S$
and D$^0$(2308) from $6_A$ -- $\bar{3}_A$ mixing. Because of that
mixing, their flavor wave functions are given only in a first
approximation \cite{jaff77a} and therefore calculations are not
sufficiently precise. Therefore these two states have theoretical
predictions which are not the same as the experimental ones. But
nevertheless, for meson fit, their experimental masses are still
between the two values we obtained in Table \ref{tab05}:
D$_s^+$(2632) has the mass between 2547 and 2936 MeV, while
D$^0$(2308) has the mass between 2195 and 2317 MeV. Obtained
theoretical masses for baryon fit in case of D$^0$(2308) is lower
than it it expected and for D$_s^+$(2632) is approximately in the
expected range. If we compare Tables \ref{tab05} and \ref{tab06} and
Figures \ref{fig01} and \ref{fig02} we can conclude that FB
interaction gives lower total masses. FB interaction that we applied
needs some further improvement. Probably the biggest discrepancy
between theoretically obtained and experimentally detected state is
because of sensitivity of constituent quark masses to the system in
which they are contained (meson, baryon, tetraquark).

\section{Conclusions}

We calculated the mass spectra of two-quark-two-antiquark system. We
used the observed meson masses, taken from "Particle Data Group"
\cite{PDG10}, to obtain constituent quark masses by way of $\chi^2$
mass fit. Applying the Hamiltonian (\ref{equ01}) to the constituent
quarks, we obtained the theoretical meson, baryon and single-charm
tetraquark masses with FB contribution included. For the first time
the FB HFI, which is color-spin interaction, is applied to 27 states
of $cq \bar{q} \bar{q}$ tetraquarks in order to derive formulas for
masses and these are new results. Also, when comparing with GR HFI,
which is flavor-spin interaction, it is interesting that we also
obtained similar results for tetraquark masses, like in the case of
meson and baryon masses. In our constituent quark model, FB HFI is
calibrated in mesons and baryons, and then used in tetraquarks.

If the tetraquark states we have studied contained no heavy quarks,
their mass would be too low; if they contained one heavy quark ($b$
or $t$) or two $c$-quarks, mass would be too high. According to that
and to the mass constraint of the experimentally indicated states,
these states can contain no more than one $c$-quark. We also showed
\cite{bork10} that the constituent quark masses are very sensitive
to the system in which they are contained, and their values differ
less or more in different systems. That is why we analyzed
tetraquark masses using meson and baryon fits for constituent quark
masses, independently. Constituent quark masses in tetraquarks are
somewhat different for both fits. We choose relatively simple FB HFI
because it nicely satisfy Roncaglia inequalities for mass
differences \cite{ronc95a,ronc95b}.

Symmetric $\overline{15}_S$-plet mixes with the $\bar{3}_S$-plet
ideally (D$_s$ and D states). This mixing splits the two states into
a heavy (hidden strangeness) and a light one. Also, antisymmetric
$6_A$-plet mixes with $\bar{3}_A$-plet (lowest mass D is the ideal
mixture of these antisymmetric multiplets, while the lowest D$_s$ is
pure $\bar{3}_A$-plet). From the spectra of single-charm tetraquark
masses with Fermi-Breit HFI, it can be noticed that this interaction
implies no flavor dependent splitting among multiplets.

Flavor wave functions of the mixed states are given only in a first
approximation (see Ref. \cite{jaff77a}). The mixing of the states
also changes the properties and shifts masses from the theoretical
predictions. For instance, possibly tetraquark states D$^0$(2308)
and D$_s^+$(2632) in our case are mixed states and the calculations
of their masses are not sufficiently precise. D$_s^+$(2632) appears
as a mixed state from mixing of multiplets $\overline{15}_S$ and
$\bar{3}_S$, and D$^0$(2308) would be from mixing $6_A$ and
$\bar{3}_A$. According to our results, all three states
D$_s^+$(2632), D$^0$(2308), D$_s^+$(2317) might have the tetraquark
nature. We gave the contribution of FB HFI to $cq \bar{q} \bar{q}$
tetraquark masses and also we calculated tetraquark masses and
compared them to GR HFI.

As it can be seen from Tables \ref{tab05} and \ref{tab06} and Figs.
\ref{fig01} and \ref{fig02}, FB HFI reduces the obtained masses and
causes splitting between $\Sigma$ and $\Sigma_s$ in
$\overline{15}_S$ and between $\Sigma_s$ and $\Omega$ in $6_A$.
Besides, the spectra obtained from different fits have a similar
arrangement of particles. Probably the biggest difference between
theoretical and experimental states is due to:
\newline
(i) sensitivity of constituent quark masses on systems in which they
are contained,
\newline
(ii) wave functions for the two detected experimental states
D$^0$(2308) and D$_s^+$(2632) are calculated only in a first
approximation, and obtained mases are not precise,
\newline
(iii) FB HFI is not the completed HFI.

More experimental searches for detection of other $cq \bar{q}
\bar{q}$ members are needed in the future.

\begin{acknowledgement}
This research is part of the project 176003 ''Gravitation and the
large scale structure of the universe'' supported by the Ministry of
Education and Science of the Republic of Serbia.
\end{acknowledgement}


\begin{thebibliography}{99}

\bibitem{jaff77a} R. L. Jaffe, Phys. Rev. D \textbf{15}, 267 (1977).

\bibitem{jaff77b} R. L. Jaffe, Phys. Rev. D \textbf{15}, 281 (1977).

\bibitem{silv92} B. Silvestre-Brac, Phys. Rev. D \textbf{46}, 2179
    (1992).

\bibitem{aube03} BABAR Collab. (B. Aubert \emph{et al}.) Phys. Rev.
    Lett. \textbf{90}, 242001 (2003).

\bibitem{mika04} BELLE Collab. (Y. Mikami \emph{et al}.) Phys. Rev.
    Lett. \textbf{92}, 012002 (2004).

\bibitem{abe04} BELLE Collab. (K. Abe et \emph{al}.) Phys. Rev. D
    \textbf{69}, 112002 (2004).

\bibitem{link04} FOCUS Collab. (J. M. Link \emph{et al}.) Phys.
    Lett. B \textbf{586}, 11 (2004).

\bibitem{evdo04} SELEX Collab. (A. V. Evdokimov \emph{et al.}) Phys.
    Rev. Lett. \textbf{93}, 242001 (2004).

\bibitem{beve03} E. van Beveren, G. Rupp, Phys. Rev. Lett.
    \textbf{91}, 012003 (2003).

\bibitem{beve04} E. van Beveren, G. Rupp, Phys. Rev.
    Lett. \textbf{93}, 202001 (2004).

\bibitem{barn04} T. Barnes, F. E. Close, J. J. Dudek, S. Godfrey, S.
    E. Swanson, Phys. Lett. B \textbf{600}, 223 (2004).

\bibitem{bard03} W. A. Bardeen, E. J. Eichten, C. T. Hill, Phys.
    Rev. D \textbf{68}, 054024 (2003).

\bibitem{wang06} Z. G. Wang, S L Wan, Nucl. Phys. A \textbf{778}, 22
    (2006).

\bibitem{wang07} Z. G. Wang, Phys. Rev. D \textbf{75}, 034013
    (2007).

\bibitem{swan06} E. S. Swanson, Phys. Rep. \textbf{429}, 243 (2006).

\bibitem{molo09} S. V. Molodtsov, T. Siemiarczuk, A. N. Sissakian,
    A. S. Sorin, G. M. Zinovjev, Eur. Phys. J. C \textbf{61}, 61
    (2009).

\bibitem{haya06} A. Hayashigaki, K. Terasaki, Prog. Theor. Phys.
    \textbf{114}, 1191 (2006).

\bibitem{tera06} K. Terasaki, Prog. Theor. Phys. \textbf{116} 435
    (2006).

\bibitem{tera05} K. Terasaki, B. H. J. McKellar, Prog. Theor. Phys.
    \textbf{114}, 205 (2005).

\bibitem{dmit05} V. Dmitra\v{s}inovi\'{c}, Phys. Rev. Lett.
    \textbf{94}, 162002 (2005).

\bibitem{niel06} M. Nielsen, R. D. Matheus, F. S. Navarra, E. M.
    Bracco, A. Lozea, Nucl. Phys. B (Proc. Suppl.) \textbf{161}, 193
    (2006).

\bibitem{nico04} B. Nicolescu, J. P. B. C. de Melo,
    arXiv:hep-ph/0407088 (2004).

\bibitem{liu04} Y.-R. Liu, S.-L. Zhu, Y.-B. Dai, C. Liu,
    Phys. Rev. D \textbf{70}, 094009 (2004).

\bibitem{brac05} M. E. Bracco, A. Lozea, R. D. Matheus, F. S.
    Navarra, M. Nielsen, Phys. Lett. B \textbf{624}, 217 (2005).

\bibitem{brit05} T. V. Brito, F. S. Navarra, M. Nielsen, E. M.
    Bracco, Phys. Lett. B \textbf{608}, 69 (2005).

\bibitem{thoo08} G. 't Hooft, G. Isidori, L. Maiani, A. D. Polosa,
    V. Riquer,  Phys. Lett. B \textbf{662}, 424 (2008).

\bibitem{vija06} J. Vijande, F. Fern\'{a}ndez, A. Valcarce, Phys.
    Rev. D \textbf{73}, 034002 (2006).

\bibitem{vija07} J. Vijande, F. Fern\'{a}ndez, A. Valcarce, Eur.
    Phys. J. A \textbf{31}, 722 (2007).

\bibitem{eber08} D. Ebert, R. N. Faustov, V. O. Galkin,
    Eur. Phys. J. C     \textbf{58}, 399 (2008).

\bibitem{zhu08} S.-L. Zhu, Nucl. Phys. A \textbf{805}, 221c (2008).

\bibitem{wang09} Z. G. Wang,  Phys. Rev. D \textbf{79}, 094027
    (2009).

\bibitem{ruju75} A. De R\'{u}jula, H. Georgi, S. L. Glashow, Phys.
    Rev. D \textbf{12}, 147 (1975).

\bibitem{libe77} D. A. Liberman, Phys. Rev. D \textbf{16}, 1542
    (1977).

\bibitem{fran81} J. Franklin, D. B. Lichtenberg, W. Namgung, D.
    Carydas, Phys. Rev. D \textbf{24}, 2910 (1981).

\bibitem{luch91} W. Lucha, F. Sch\"{o}berl, D. Gromes, Phys. Rep.
    \textbf{200}, 127 (1991).

\bibitem{bork10} V. Borka Jovanovi\'{c}, S. R. Ignjatovi\'{c}, D.
    Borka, P. Jovanovi\'{c}, Phys. Rev. D \textbf{82}, 117501
    (2010).

\bibitem{ronc95a} R. Roncaglia, A. Dzierba, D. B. Lichtenberg, E.
    Predazzi, Phys. Rev. D \textbf{51}, 1248 (1995).

\bibitem{ronc95b} R. Roncaglia, D. B. Lichtenberg, E. Predazzi,
    Phys. Rev. D \textbf{52}, 1722 (1995).

\bibitem{giam95} J. Giammarco, J. Franklin,  Nucl. Phys. A
    \textbf{585}, 450 (1995).

\bibitem{gloz96} L. Ya. Glozman, D. O. Riska, Phys. Rep.
    \textbf{268}, 263 (1996).

\bibitem{bork07} V. Borka Jovanovi\'{c}, Phys. Rev. D \textbf{76},
    105011 (2007).

\bibitem{bork08} V. Borka Jovanovi\'{c}, Fortschr. Phys.
    \textbf{56}, 462 (2008).

\bibitem{yasu07} S. Yasui, M. Oka, Phys. Rev. D \textbf{76}, 034009
    (2007).

\bibitem{feng08a} X.-C. Feng, Acta Phys. Polonica B \textbf{39},
    2931 (2008).

\bibitem{feng08b} X.-C. Feng, F.-C. Jiang, T.-Q. Chang, J.-L. Feng,
    Chinese Phys. B \textbf{17}, 4472 (2008).

\bibitem{PDG10} K. Nakamura, et al. (Particle Data Group) J. Phys.
    G \textbf{37}, 075021 [http://pdg.lbl.gov] (2010).

\bibitem{maia05} L. Maiani, F. Piccinini, A. D. Polosa, V. Riquer,
    Phys. Rev. D \textbf{71}, 014028 (2005).

\end{thebibliography}
\end{document}